\documentclass[10pt]{article}
\usepackage{graphicx}

\def\Title#1{\begin{center} {\Large #1 } \end{center}}
\def\Author#1{\begin{center}{ \sc #1} \end{center}}
\def\Address#1{\begin{center}{ \it #1} \end{center}}

\newcommand\pubblock{\rightline{\begin{tabular}{l} Proceedings of the Fifth Annual LHCP\\ \pubnumber\\
         \pubdate  \end{tabular}}}

\newenvironment{Abstract}{\begin{quotation} \begin{center} 
             \large ABSTRACT \end{center}\bigskip 
      \begin{center}\begin{large}}{\end{large}\end{center} \end{quotation}}

\newenvironment{Presented}{\begin{quotation} \begin{center} 
             PRESENTED AT\end{center}\bigskip 
      \begin{center}\begin{large}}{\end{large}\end{center} \end{quotation}}

\def\beq{\begin{equation}}
\def\eeq#1{\label{#1}\end{equation}}
\def\eeqn{\end{equation}}

\def\beqa{\begin{eqnarray}}
\def\eeqa#1{\label{#1}\end{eqnarray}}
\def\eeqan{\end{eqnarray}}

\let\bar=\overbar

\def\Dslash{\not{\hbox{\kern-4pt $D$}}}
\def\dslash{\not{\hbox{\kern-2pt $\del$}}}

\def\msb{{\bar{\ssstyle M \kern -1pt S}}}

\usepackage{color}

\newcommand\pubnumber{ CMS-CR-2017-201 }

\newcommand\pubdate{\today}

\def\affiliation{
On behalf of the CMS Experiment, \\
University of Wisonsin-Madison, Madison, United States of America}

\begin{document}

\large
\begin{titlepage}
\pubblock

\vfill
\Title{Search for exotic decays of the Higgs boson}
\vfill

\Author{ Cecile Caillol }
\Address{\affiliation}
\vfill
\begin{Abstract}
Searches for exotic decays of the 125 GeV Higgs boson performed with data collected by the CMS experiment are presented. Three classes of searches are detailed: searches for invisible decays of the Higgs boson, searches for lepton flavor violating Higgs decays, and searches for decays to the Higgs boson to light pseudoscalars decaying to SM particle pairs. These analyses are based on data collected at center-of-mass energies of 8 and 13 TeV.

\end{Abstract}
\vfill

\begin{Presented}
The Fifth Annual Conference\\
 on Large Hadron Collider Physics \\
Shanghai Jiao Tong University, Shanghai, China\\ 
May 15-20, 2017
\end{Presented}
\vfill
\end{titlepage}
\def\thefootnote{\fnsymbol{footnote}}
\setcounter{footnote}{0}
%

\normalsize 


\section{Introduction}

After the recent discovery of the Higgs boson in 2012 by the ATLAS and CMS experiments~\cite{Aad:2012tfa,Chatrchyan:2012ufa}, many measurements have confirmed that the new particle with mass of 125 GeV has properties, including spin, CP, and coupling strengths to  standard model (SM) particles, that are compatible with those expected for the Higgs boson of the SM. However the scalar sector is not well known experimentally yet, and current measurements could still accommodate for large contributions of new physics in this sector. Observing exotic decays of the Higgs boson would be a striking direct evidence for the existence of physics beyond the SM (BSM).

\section{Invisible decays}

Higgs bosons can decay to invisible particles in various BSM models, including in supersymmetric models with decays to pairs of lightest stable particles (LSP), or models with large extra dimensions with mixing of graviscalars with the Higgs boson. Different signatures can be studied depending on the production mode of the Higgs boson.\\

The ZH production mode, with the Z boson subsequently decaying to a dimuon or a dielectron pair, offers a clean signature. The analysis looks for a Z boson candidate recoiling against a large missing transverse momentum~\cite{EXO-16-052}. The Z boson candidate is formed from two well-isolated same-flavor opposite-sign leptons with a mass compatible with the Z boson mass. Results are extracted from a binned likelihood fit to the missing transverse momentum distribution. Using as the observable the output of a dedicated boosted decision tree (BDT) improves the sensitivity of the analysis by about 10\%. No excess of events is observed on top of the background prediction. Observed (expected) upper limits at 95\% confidence level equal to 40\% (42\%) are set on the branching fraction of the Higgs boson to invisible particles, using 35.9 fb$^{-1}$ of proton-proton data collected by the CMS experiment at 13 TeV center-of-mass energy.\\

Events with Higgs bosons decaying invisibly and produced via vector boson fusion are characterized by the presence of two forward jets with a large invariant mass. At trigger level, two jets with $p_T>40$ with a pseudorapidity difference greater than 3.5 and an invariant mass greater than 600 GeV are required~\cite{HIG-16-009}. The QCD multijet background is reduced by requiring the jets to recoil against a large missing transverse momentum. Control regions with one or two leptons are added to the fit to constrain the yields of the major backgrounds. Observed (expected) upper limits at 95\% confidence level equal to 69\% (62\%) are set on the branching fraction of the Higgs boson to invisible particles, using 2.8 fb$^{-1}$ of proton-proton data collected by the CMS experiment at 13 TeV center-of-mass energy.\\

Events with 1 jet or 2 jets recoiling against a large transverse missing momentum are the signature of Higgs bosons produced in gluon fusion with one jet from initial state radiation, or in VH production with hadronic V boson decays, respectively~\cite{EXO-16-048}. Various control regions with leptons in the final state are included in the fit to extract the results, together with the missing transverse momentum distributions in the signal region. The combination of both analyses with 35.9 fb$^{-1}$ of proton-proton data collected by the CMS experiment at 13 TeV center-of-mass energy lead to observed (expected) upper limits at 95\% confidence level equal to 53\% (40\%) are set on the branching fraction of the Higgs boson to invisible particles.\\

\begin{figure}[htb]
\centering
\includegraphics[height=3in]{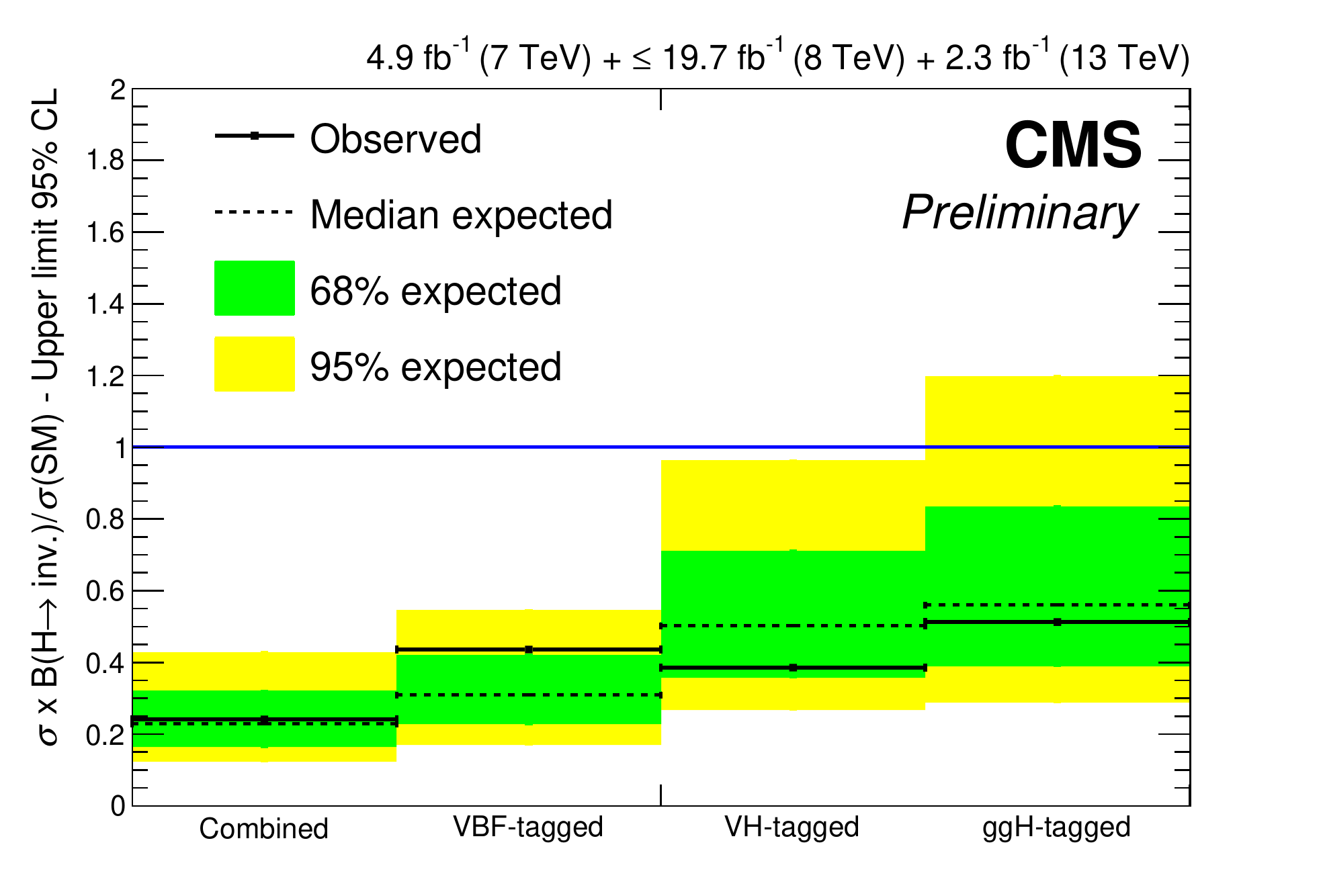}
\caption{ Observed and expected 95\% CL limits on $(\sigma_h/\sigma_{SM})\mathcal{B}(H\rightarrow\textrm{inv})$ for individual combinations of categories targeting VBF, VH, and ggH production, and the full combination assuming a Higgs boson with the mass of 125 GeV.}
\label{fig:figure1}
\end{figure}

The combination of searches for invisible Higgs boson decays in all the production modes~\cite{Hinv}, using data collected at 7, 8, and 13 TeV center-of-mass energies (where 13 TeV data correspond to 2.8 fb$^{-1}$ collected in 2015) leads to observed (expected) upper limits at 95\% confidence level equal to 24\% (23\%) are set on the branching fraction of the Higgs boson to invisible particles, as shown in Fig.~\ref{fig:figure1}.

\section{Lepton flavor violating decays}

Searches for $H\rightarrow e\tau$ and $H\rightarrow\mu\tau$ decays have been performed with 35.9 fb$^{-1}$ of ptoton-proton data collected by the CMS experiment at 13 TeV center-of-mass energy~\cite{HIG-17-001}. Final states with a hadronic $\tau$, denoted $\tau_h$, and with two different-flavor light leptons are considered. Same-flavor light lepton final states are discarded because of the large background from Drell-Yan production.\\

The events are separated into four different categories, with various signal-to-background ratios:
\begin{itemize}
\item 0 jet. The events do not contain any jet with $p_T>30$ GeV and $|\eta|<4.7$. This category targets Higgs bosons produced via gluon fusion.
\item 1 jet. The events contain one jet with $p_T>30$ GeV and $|\eta|<4.7$. This category targets Higgs bosons produced via gluon fusion with one jet from initial state radiation.
\item 2 jets, gg-enriched. The events contain two jets with $p_T>30$ GeV and $|\eta|<4.7$. The invariant mass of the two jets is required to be less than 500 or 550 GeV. This category targets Higgs bosons produced via gluon fusion with several jets from initial state radiation.
\item 2 jets, VBF-enriched. The events contain two jets with $p_T>30$ GeV and $|\eta|<4.7$. The invariant mass of the two jets is required to be larger than 500 or 550 GeV. This category targets Higgs bosons produced via vector boson fusion.
\end{itemize}

The background from Drell-Yan production is estimated from simulations. Corrections to the di-lepton mass and di-lepton $p_T$ are extracted from a $Z\rightarrow\mu\mu$ control region, and applied to the Drell-Yan background in the signal region. 
The $t\bar{t}$, diboson, single top quark, and SM Higgs boson production backgrounds are also estimated from simulations and normalized to their predicted theoretical cross sections. The W+jets and QCD multijet backgrounds, where at least one jet is misidentified as a final state lepton, are partially estimated from data.\\

The main handles to discriminate the signal from backgrounds are:
\begin{itemize} 
\item The transverse missing momentum is close to be aligned with the visible $\tau$ decay product because neutrinos from $\tau$ decays are the only source of transverse missing momentum.
\item Signal events typically have higher momentum muons and electrons, in comparison for example to electrons and muons from $\tau$ decays in $Z\rightarrow\tau\tau$ events.
\item The visible invariant mass of the signal leptons forms a peak at a different position than the narrow peak from $Z\rightarrow ee/\mu\mu$ or than the wide peak of $Z\rightarrow\tau\tau$ decays. It is also possible to determine an unbiased estimation of the mass of the Higgs boson by using the collinear approximation, which assumes that the neutrinos from $\tau$ decays are aligned with the visible $\tau$ decay products.
\end{itemize}
Variables related to these event characteristics are combined in a boosted decision tree (BDT) for each channel. The BDT is trained against reducible background events for the $e\tau_h$ and $\mu\tau_h$ final states, whereas it is trained against the $t\bar{t}$ (or $t\bar{t}$ + diboson production) background in the $e\tau_\mu$ and $\mu\tau_e$ final states. \\

The results are extracted with a binned likelihood fit of the BDT output distributions in the various channels and categories. No significant excess of events is observed on top of the background expectation. Upper limits at 95\% confidence level are set on the branching fraction of the Higgs boson to $e\tau$ and $\mu\tau$. The observed and expected limits are both 0.25\% for the $\mu\tau$  decay channel, and 0.63\% and 0.37\% for the $e\tau$ decay channel, respectively. The results are cross-checked by selecting the events with selection criteria on different variables and extracting the results with a fit the collinear mass distributions. The results are compatible, with a lower sensitivity.

\begin{figure}[htb]
\centering
\includegraphics[height=3.5in]{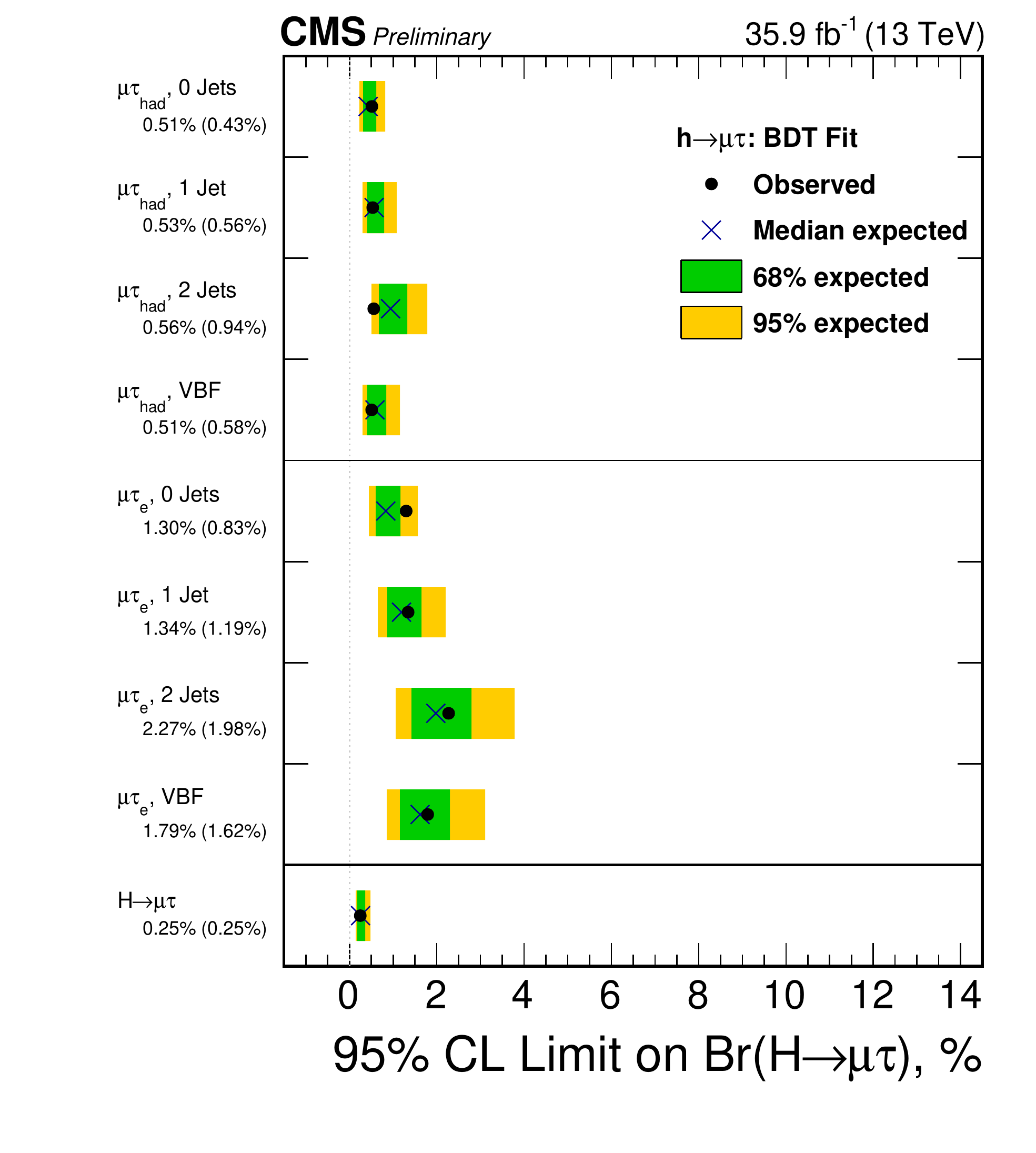}
\includegraphics[height=3.5in]{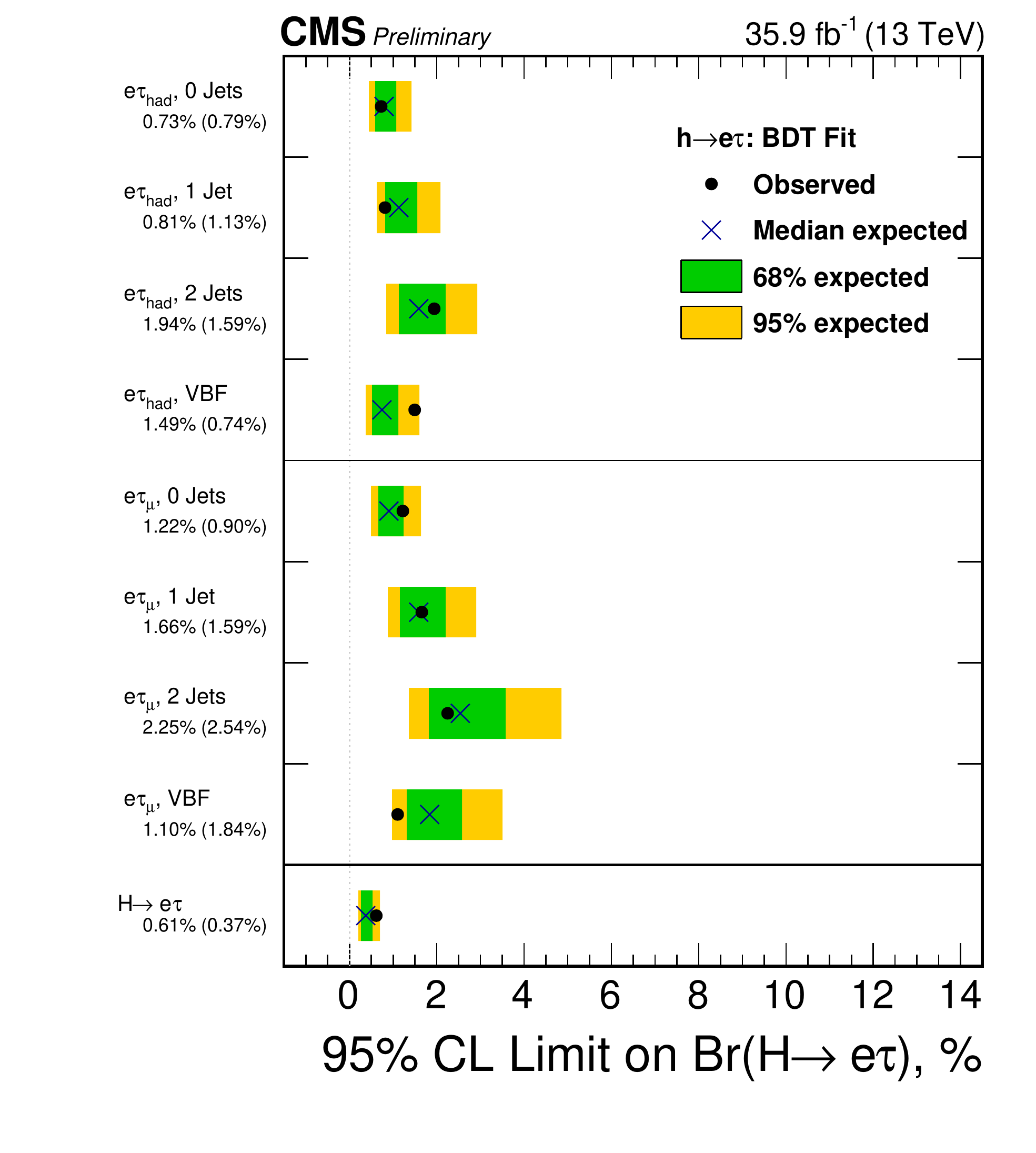}
\caption{ Observed and expected 95\% CL limits on $\mathcal{B}(H\rightarrow\mu\tau)$ (left) and $\mathcal{B}(H\rightarrow e\tau)$ (right).}
\label{fig:figure2}
\end{figure}

\section{Decays to light pseudoscalars}

\begin{figure}[htb]
\centering
\includegraphics[height=3in]{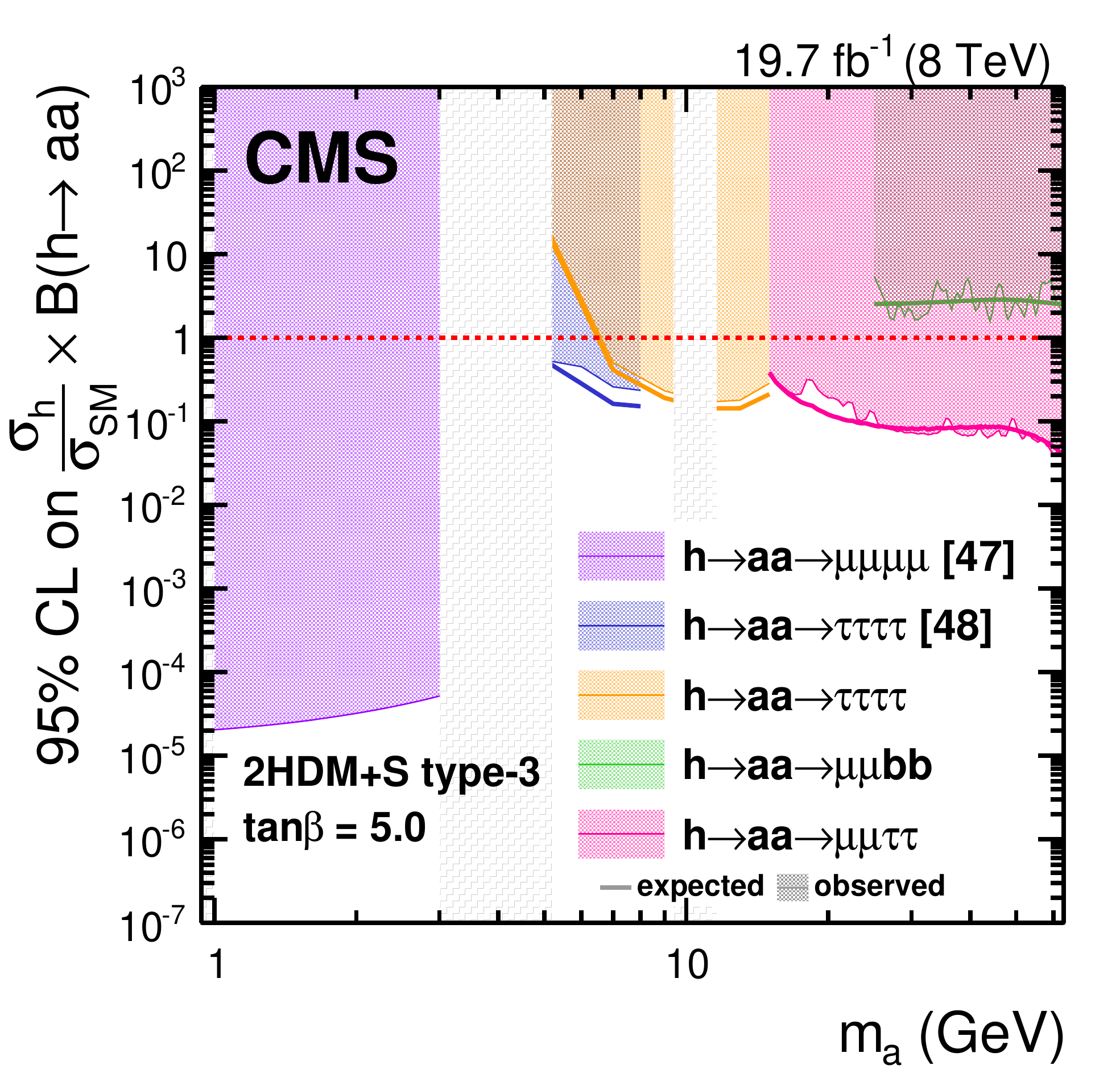}
\caption{ Expected and observed 95\% CL limits on $(\sigma_h/\sigma_{SM})\mathcal{B}(h\rightarrow aa)$ in 2HDM+S type-3 with $\tan\beta =5$. Limits are shown as a function of the mass of the light boson, $m_a$. Grey shaded regions correspond to regions where theoretical predictions for the branching fractions of the pseudoscalar boson to SM particles are not reliable.}
\label{fig:figure3}
\end{figure}

Models with two scalar doublets and one scalar singlet, also called 2HDM+S, predict the existence of 7 scalars and pseudoscalars. Among them, one scalar, $h$, can be compatible with the 125 GeV Higgs boson, and another one, $a$, can be a light pseudoscalar such that $h\rightarrow aa$ decays are allowed. Such decays are allowed after all LHC measurements, and the branching fraction for $h\rightarrow aa$ can be as large as about 20\%. Five searches for $h\rightarrow aa$ decays have been performed with data collected by the CMS experiment at center-of-mass energy of 8 TeV~\cite{HIG-16-015}. The results of these searches are interpreted in various 2HDM+S scenarios. The interpretation in type-3 2HDM+S (characterized by an increased couplings to the pseudoscalar boson to leptons at large $\tan\beta$, defined as the ratio of the vacuum expectation values of the two doublets) with $\tan\beta=5$, is shown in Fig.~\ref{fig:figure3}.\\

The $h\rightarrow aa\rightarrow 4\mu$ analysis is performed for masses of the pseudoscalar boson between 0.25 and 3.5 GeV, before the $a\rightarrow2\tau$ decay channel opens. This analysis~\cite{HIG-16-035} benefits from the excellent dimuon mass resolution. The dominant background arises from the $b\bar{b}$ production. The background templates are extracted from data, and parameterized with a sum of Bernstein polynomials, Gaussians, and Crystal Ball functions to fit all SM resonances ($\omega$, J/$\psi$, $\phi$, $\psi$). In the signal region, where the mass of the dimuon pairs is required to be compatible with each other within detector resolution effects, one event is observed, in agreement with the background expectation of $0.74\pm0.74$(stat.)$\pm 0.15$(syst.), using 2.8 fb$^{-1}$ of data collected in 2015 at center-of-mass energy of 13 TeV.\\

Two analyses focus on $h\rightarrow aa\rightarrow 4\tau$ decays for pseudoscalar masses between 5 and 15 GeV, with different technical solutions. In this mass region the ditau pairs are boosted, requiring special techniques to reconstruct the $\tau$ leptons. The first analysis requires the presence of a high-$p_T$ muon, used to trigger the event, and of a boosted $\tau$ pair composed of a muon (from $\tau$ decay) and of a nearby object passing the $\tau_h$ reconstruction and isolation conditions when the muon is removed from its isolation cone~\cite{HIG-16-015}. A category with large transverse mass between the first muon and the missing transverse momentum is designed to specifically target the WH associated production, with muonic decay of the W boson. The second analysis~\cite{H4tau} reconstructs boosted $\tau$ pairs from a muon and a nearby opposite-sign track. Both analyses have a comparable sensitivity and cover different mass ranges.\\

The $h\rightarrow aa\rightarrow 2\mu 2\tau$ channel benefits from a clean final state and a narrow dimuon mass resonance, but suffers from relatively small branching fractions in most models. The analysis is performed for pseudoscalar masses between 15 and 62.5 GeV, leading to well separated leptons~\cite{HIG-16-015}. All possible $\tau\tau$ final states are considered except for the fully muonic one  because of complications in the pair reconstruction with the other two muons. The invariant mass of the four leptons is required to be compatible with 125 GeV, and the results are extracted by a fit to the dimuon mass distributions. The signal distribution is parameterized with a Voigtian or a Lorentzian profile, the irreducible ZZ$\rightarrow 4\ell$ with a fifth-degree Bernstein polynomial, and the reducible background (mostly Z+jets and WZ+jets, estimated from data with the fake rate method) with a third-degree polynomial. \\

For pseudoscalar masses between 25 and 62.5 GeV, the $h\rightarrow aa\rightarrow 2\mu 2b$ channel is sensitive channel because of the narrow dimuon mass peak and because of large branching fractions in many models allowing decays of the pseudoscalar to quarks~\cite{HIG-16-015}. The results are extracted from an unbinned fit to the dimuon mass distribution, after requiring the mass of the four final state objects to be compatible with 125 GeV.\\

No significant excess of events is observed above the background expectation for any of these searches.

\section{Summary}

Searches for exotic decays of the 125 GeV Higgs boson performed with data collected with the CMS experiment have been presented. No excess of data above the predicted backgrounds has been seen in any search for invisible Higgs boson decays, lepton-flavor violating Higgs boson decays, nor Higgs boson decays to pseudoscalar pairs. Stringent limits have been set on the existence of such processes.

\end{document}